\documentclass[preprint,showpacs,showkeys,preprintnumbers,amsmath,amssymb,superscriptaddress]{revtex4}

\usepackage{graphicx}
\usepackage{dcolumn}
\usepackage{bm}

\begin{document}


\title{Numerical Study of the Oscillatory Convergence to the Attractor 
at the Edge of Chaos.}

\author{R. Tonelli}
 \email{roberto.tonelli@dsf.unica.it}
 \affiliation{Physics Department and INFN, University of Cagliari,  Italy.}
 \affiliation{SLACS Laboratory.}
\author{M. Coraddu}%
 \email{massimo.coraddu@ca.infn.it}
\affiliation{Physics Department and INFN, University of Cagliari,
 Italy.}

\date{\today}

\begin{abstract}
This paper compares three different types of ``onset of chaos''
in the logistic and generalized logistic map: the Feigenbaum attractor
at the end of the period doubling bifurcations; the tangent bifurcation at the
border of the period three window; the transition to chaos in the generalized logistic
with inflection 1/2 ($x_{n+1} = \mu x_{n}^{1/2}$), in which
the main bifurcation cascade, as well as
the bifurcations generated by the periodic windows in the chaotic region,
collapse in a single point.
The occupation number and the Tsallis entropy are studied.
The different regimes of convergence to the attractor,
starting from two kinds of far-from-equilibrium initial conditions,
are distinguished by the presence or absence of log-log
oscillations, by different power-law scalings and by a gap in the
saturation levels.
We show that the escort distribution 
implicit in the Tsallis entropy may tune 
the log-log oscillations or the crossover times.

\end{abstract}

\maketitle

\section{Introduction}

The transition from regular to chaotic regime presents characteristics similar 
to phase transitions in statistical thermodynamics and an adequate statistical 
thermo-dynamical formalism has been developed for chaotic systems (see ref. 
\cite{beck}
for an introduction).
The transition point, or border of chaos, is characterized by a null Lyapunov exponent. 
The interesting features of this transition were first illustrated 
by the pioneering works of Feigenbaum on the logistic map attractor at the infinite 
bifurcation point
\cite{feig}, immediately followed by a series of theoretical works 
\cite{grass81}, \cite{politi}, \cite{mori}. In these investigations 
a power-law behavior of the sensitivity function of the logistic map at the 
threshold of chaos was formulated and the values for the power-law exponent 
were analytically calculated. 
More recently Tsallis \cite{Ts:88} introduced a formalism of non-extensive 
thermodynamics that allows to pass easily from the chaotic case to the 
null Lyapunov exponent case, recovering the Boltzmann Gibbs (BG) formulation 
as a limit. This formalism enlightens the connections between chaos and 
border of chaos and, in particular, defines a generalized entropy as the quantity of 
physical interest. For example it has been shown 
\cite{An:04} that a strong analogy 
with the Pesin Identity \cite{Pesin} exists among this generalized entropy 
and generalized polynomial sensitivity to initial conditions appropriate 
to describe the sensitivity at 
the infinite bifurcations point of the logistic map \cite{Ba:02}. 

Also other statistical formalisms emerging within special relativity 
\cite{kaniadakis} and
quantum groups \cite{abe}
seem to 
describe well 
\cite{noi} the analogies between the threshold of chaos and the fully chaotic regime, 
even if recently the non-extensive formalism has been strongly criticized 
\cite{grass2005} revealing a still open debate on the subject. 
At this time a lot of experimental results (see \cite{Ts:rev} for a review) 
seem to confirm at least part of the theoretical framework of the
non-extensive formalism.
In this paper we performed various 
numerical experiments devoted to investigate the logistic map 
at transition points presenting different routes to chaos. 
Thus our study regards different border of chaos regimes 
with different attractors underlying the dynamics. 
We present a joint analysis of (generalized, coarse grained) entropy and 
occupation number. 
We investigate the dynamics at large times and at large sampling ratios 
and analyze border of chaos transitions different from
the Feigenbaum attractor in the logistic map. 
Finally we examine the presence of oscillations
in the convergence to the attractor.

\section{Numerical experiments on the Feigenbaum attractor}

In this paper we describe a set of numerical experiments, 
performed on the logistic map at the onset of chaos, with the purpose of 
examining the dynamics of a statistical ensemble of points, 
starting from far-from-equilibrium initial conditions.  
In each experiment the available phase space is partitioned 
in a number W$_{box}$ of elementary, equal cells. 
A set of N points is randomly selected in the phase space according to 
two kinds of initial set-ups, and 
iterated according to the map. 
In these experiments we used the logistic map in the form 
\begin{equation}
x_{i+1} = 1 - \mu x_{i}^{2}\;\; ; \;\;\;\; -1 \leq x \leq 1 \;\; ; \;\;\;\; 0 \leq \mu \leq 2~.
\end{equation}
We investigated the parameter values $\mu$ = 2 for the fully chaotic case, 
$\mu_{\infty}$ = 1.401155189.. at the Feigenbaum attractor, 
$\mu_{tg}$ = 1.75 at the tangent 
bifurcation and $\mu_{1/2}$ = 2/$\sqrt{3}$ when the logistic is 
generalized with inflection 
1/2 ($x_{i+1} = 1 - \mu x_{i}^{1/2}$). 
The first kind of initial start-up is from concentrated initial conditions (i. c.). 
All the N points are chosen, with a uniform random distribution, inside 
a single cell of the partition, itself chosen at random. 
In the second kind of experiment all the N initial points are 
uniformly and randomly distributed in all the available phase space (spread 
i. c.). 
The i-th cell will contain 
a fraction N$_{i}$/N of the total number of points ($\sum_{i=1}^{W_{box}}N_{i} = N$),
so that one can naively define a probability of occupation for the i-th 
cell through $p_{i} = N_{i}/N$ with the constraint $\sum_{i=1}^{W_{box}}p_{i} = 1$. 
On one hand we observe, for both kinds of experiments (concentrated and spread i. c.)
the occupation number in time, namely the number of non 
empty cells of the partition. 
On the other hand, 
the other quantity of interest in our discussion is the physical entropy, 
defined as coarse-grained entropy through the probabilities $p_{i}$. 
We will use the Tsallis definition of entropy \cite{Ts:88}, 
useful to characterize the behavior at the onset of chaos where the Lyapunov 
exponent vanishes: 
\begin{equation}
S_{q}^{TS} = \frac{ 1 - \sum_{i=1}^{W_{box}} p_{i}^{q} }{q - 1}
\label{eq:TsEnt}
\end{equation}
Other definitions of entropies \cite{kaniadakis,abe,noi} would 
also work for the purposes of this paper. 
The analysis is performed through comparisons of the 
various behaviors between chaotic regime and border of chaos. 
We use 
different partitions W$_{box}$ and different {\em sampling ratios r}, defined
as $r$=N/W$_{box}$, that gives an indication of the goodness of sampling. 

The fact that the Lyapunov exponent is zero at the threshold of chaos, 
allows us to use power-laws to characterize the evolution \cite{grass81, politi, 
Ts:97, Ly:98}. The comparison with the chaotic case makes natural to pass 
from an "exponential formalism" to an "extended exponential formalism"
describing the power-law \cite{An:04, Ka:04}. 
As discussed in \cite{La:99} , in the Chaotic regime,
the (averaged) time evolution of the map, starting from concentrated initial conditions, 
may happen in two or three stages:
a first ``thermalization'' stage heavily dependent on the 
details of the initial conditions; a second 
``linear'' stage in which the coarse grained Boltzmann-Gibbs (BG) entropy 
$S_{BG}$\/ grows linearly with time
and a third ``saturation'' stage in which $S_{BG}$\/
reaches its final equilibrium value. 

\begin{figure}
 \begin{center}

 \includegraphics[scale=0.6]{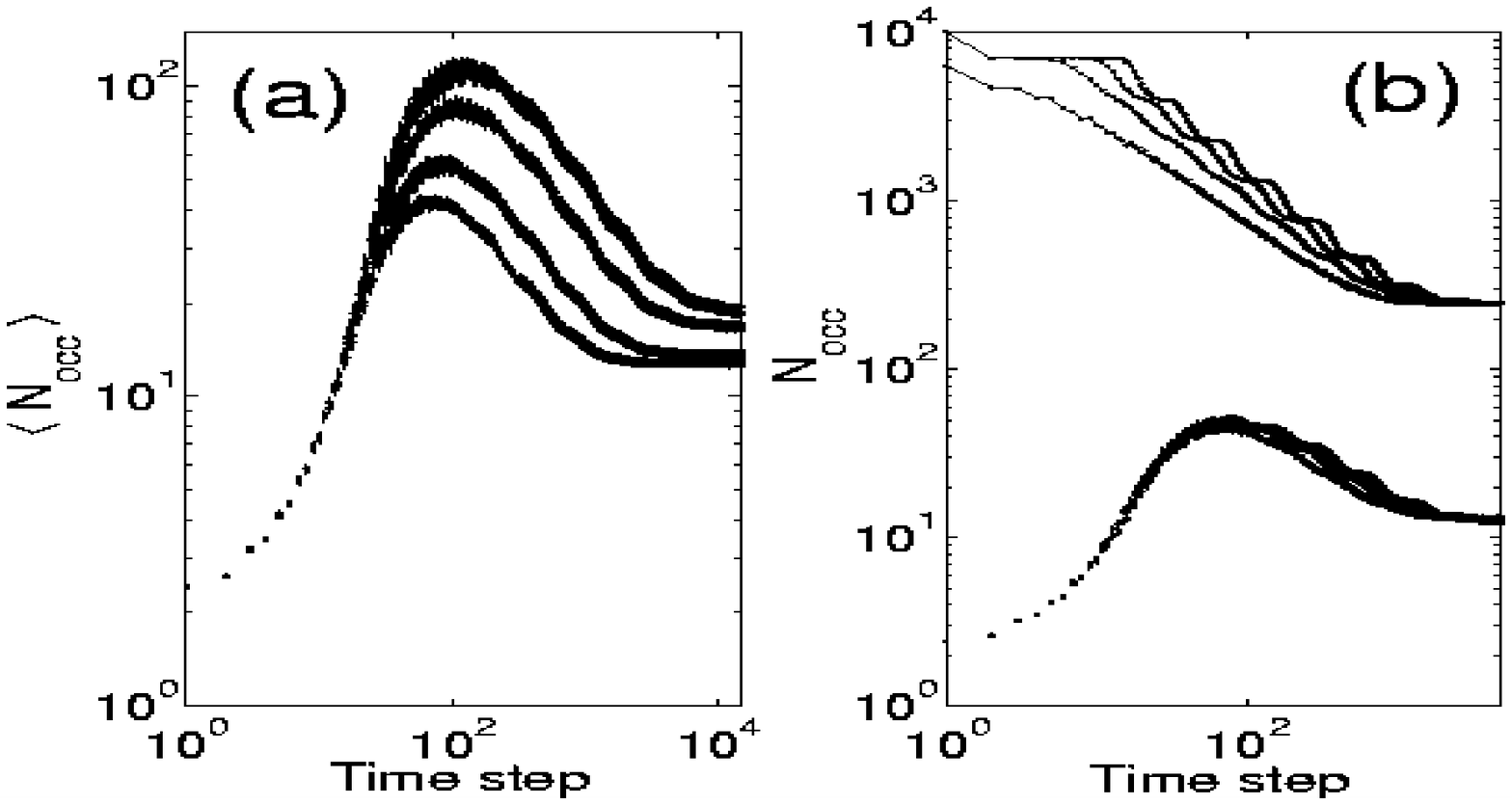}
 
 \end{center}
 \caption{\footnotesize {\bf (a)}: time evolution of the average occupation number 
 $\langle N_{occ} \rangle$
          for the Logistic map at $\mu_{\infty}$. Different curves correspond 
	  to different W$_{box}$ with $r$=1  
	  (from top to bottom 
          $W_{box}= 10^5, 5\cdot 10^4, 2\cdot 10^4, 10^4$).\\
	  {\bf (b)}: N$_{occ}$ vs. time from spread initial conditions (upper curves) 
	  and $\langle N_{occ} \rangle$ from CIC (lower curves). Fixed W$_{box}$ (10$^{4}$) and  
	  different $r$\/ (from top to bottom: $r$\/= 1000,100,10,1 upper curves, 
	  $r$\/= 100, 10, 1 
	  lower curves).
	  The averages are made over 5000 
	  randomly chosen initial configurations. }

 \label{fig:NoccCriticPoFig}
 \end{figure}

\begin{figure}
 \begin{center}

 \includegraphics[scale=0.3]{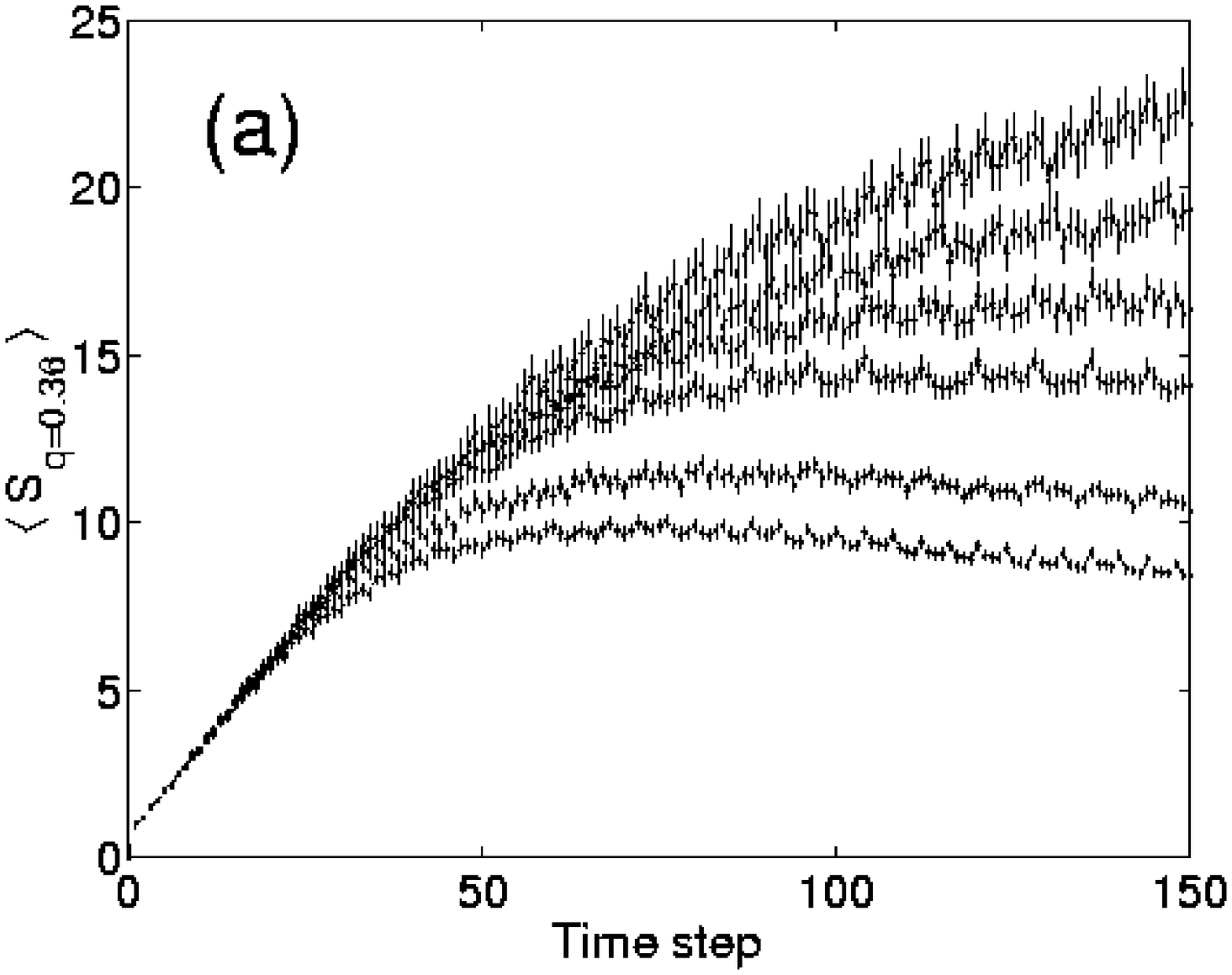}
 \includegraphics[scale=0.3]{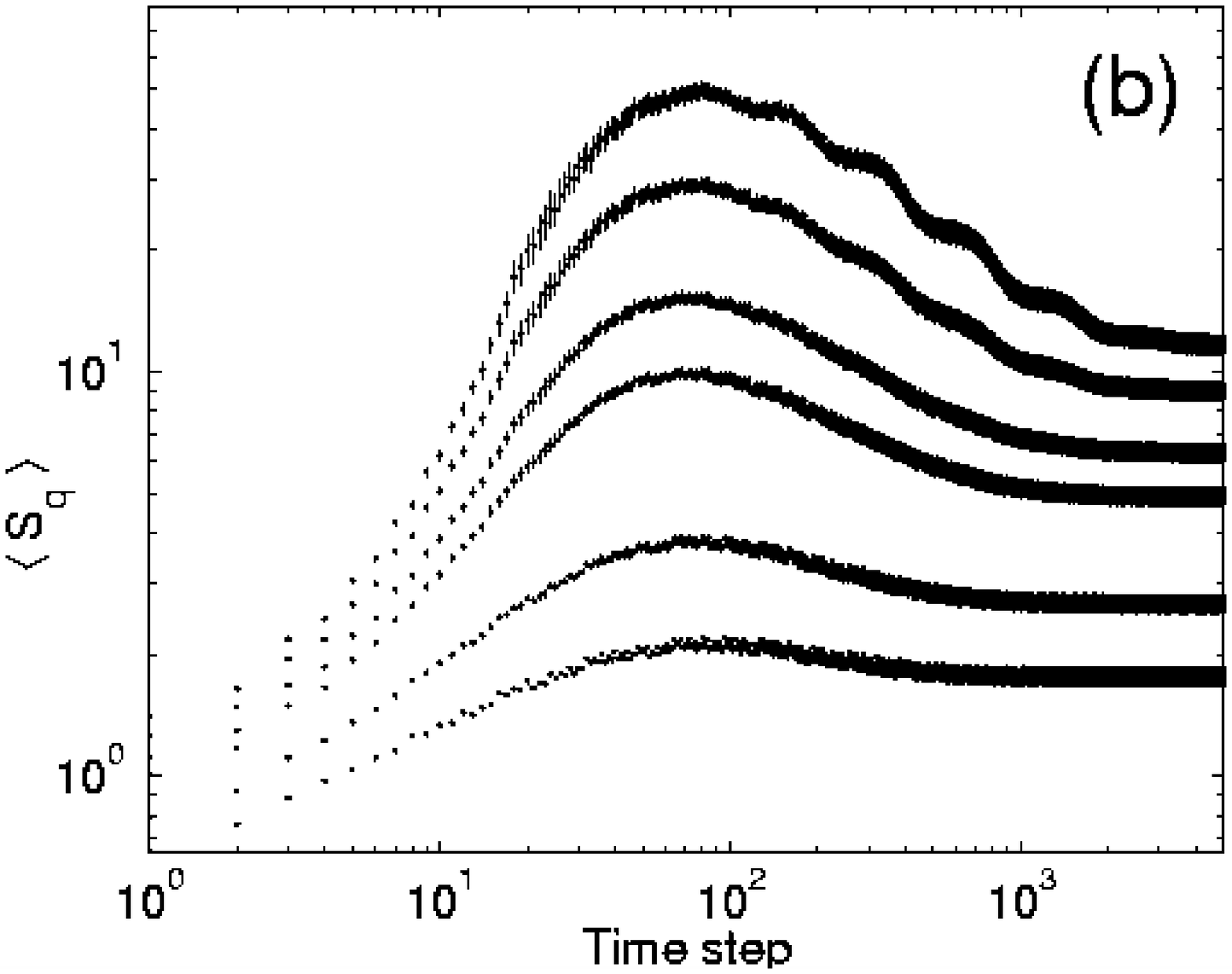}
 
 \end{center}
 \caption{\footnotesize Logistic map at the critical point
          $\mu=1.401155189$.   Tsallis Entropy  $\langle S_{q}\rangle$\/ 
          is averaged over $N_{av}=5000$\/ random choices of an expansion experiment
          initial cell and is plotted together its statistical error.
          {\bf (a)}:  different point ensembles are relative to
          increasing $W_{box}$\/ values (from top to bottom 
          $W_{box}= 5\cdot 10^5, 2\cdot 10^5, 10^5, 5\cdot 10^4, 2\cdot 10^4, 10^4$)
          while the entropic index $q=0.36$\/ (the value that linearize the first evolution stage) 
          and the sampling ratio $r=1$\/ do not vary. 
          {\bf (b)}: fixed $W_{box}=10^4$\/ and $r=100$, the point ensembles are relative to
          different entropic indexes (from top to bottom $q=0.001, 0.1, 0.2445, 0.36, 0.7, 1$).}
 \label{fig:SqCriticPoFig}
 \end{figure}

In the fully chaotic regime, 
when proper average is made on different experiments of the same kind (see ref. \cite{La:00}), 
the evolution of the logistic map, starting from concentrated initial conditions (CIC),
exhibits an initial exponential growth of the occupation number  
without a first thermalization stage. 
A saturation 
is reached, depending on the grid size, when N$_{occ}$ equals the fractal support value 
(N$_{occ}$ = W$_{box}$ for $\mu$ = 2). 
The coarse grained version of the 
BG entropy shows a linear initial 
increase followed by a saturation 
at a S$_{BG_{sat}}$\/ level
(S$_{BG_{sat}}\simeq \ln(W_{box})$)
when starting from CIC.
A Pesin-like identity 
can be observed for the coarse-grained BG entropy. 
Using for this case 
a non-extensive formalism \cite{La:00,Ba:01, An:04,noi}, 
a linear growth of S$_{q}$ in the first evolution stage can be found and 
a Pesin-like identity 
can be recovered, if the appropriate entropic index is selected . 
Thus some features of the chaotic case can be directly transferred to the 
threshold of chaos. 
It is important to note that in the chaotic case the evolution 
happens in only two stages. In this paper we are interested in clarify how 
the final saturation stage is reached at the onset of chaos. 
To this aim we studied the evolution at longer times 
than the previous works. 
The results are illustrated in Figs.\ref{fig:NoccCriticPoFig} and \ref{fig:SqCriticPoFig},
where, differently from the chaotic case, we can observe roughly three stages. 
The first stage is the analogue of the chaotic one:
both  $\langle N_{occ} \rangle$ and  $\langle S_{q} \rangle$\/ 
increase and reach a maximum. 
The duration of this first stage becomes longer and the maximum level higher 
increasing W$_{box}$, while they do not depend on $r$. 
As previously stressed \cite{An:04,noi,La:00}, in this first  ``linear'' stage,
$\langle S_{q} \rangle$\/ grows linearly only for the proper entropic index 
($q=0.36$\/ when $\mu = \mu_{\infty} $).
In the second stage both $\langle N_{occ} \rangle$ 
and $\langle S_{q} \rangle$ 
decrease.  $\langle N_{occ} \rangle$\/ follows a power-law 
with superimposed log-log oscillations  
which amplitude increases with $r$\/ and does not depend on W$_{box}$.
For low enough $q$\/ values, $\langle S_{q} \rangle$\/ 
reproduces the $\langle N_{occ} \rangle$ behavior
(when $q\simeq 0\;$\/ results $S_q\simeq N_{occ} -1$\/),
but the amplitude of oscillations decreases with increasing $q$\/ and 
vanishes when $q\sim 0.36$\/ (see fig.\ref{fig:SqCriticPoFig} right frame).  
In the third stage $\langle N_{occ} \rangle$ and  $\langle S_{q} \rangle$ 
reach their (W$_{box}$ dependent) saturation values and the evolution ends. 
Note that the saturation is reached at a level that increases with W$_{box}$,
but $\langle N_{occ} \rangle$\/ results in a 
lower final value than the fractal support, 
that is equal to $N_{occ}=244$\/ (799) for  W$_{box}=10^4$\/ ($10^5$).

The log-log oscillations of N$_{occ}$\/ 
have been already observed in the logistic map at $\mu = \mu_{\infty}$\/
\cite{Mo:00, To:05}, performing numerical experiments starting from uniform i. c. 
(see fig.\ref{fig:NoccCriticPoFig} -b). 
Here, for the first time, we can observe this same behavior 
for $\langle N_{occ} \rangle$\/ starting from  CIC
(see fig.\ref{fig:NoccCriticPoFig} -b).
This kind of logarithmic oscillations, superimposed to a power-law,
characterize a large number of systems 
exhibiting discrete scale invariance \cite{sornette}.
The time evolution $N_{occ}(t) = t^{-\delta}\, P(\ln(t))$\/
(where $P(\ln(t))$\/ is a periodic function) can be expanded in
Fourier series and, keeping only the first term, can be written in the form:
\begin{equation}
    N_{occ}(t)\, =\, t^{-\delta}\, A \left( 1\, +\, B \cos(\omega \ln(t) + \phi) \right)
\label{eq:LogLogForm}
\end{equation}
As can be observed in fig.\ref{fig:NoccCriticPoFig}-b (upper curves), 
starting from uniform i. c., 
the exponent $\delta$\/ increases with $r$\/ (this dependence has been stressed in \cite{To:05})
and its limit value ($ \delta \rightarrow 0.800138 ...$\/ for 
$r \rightarrow \infty$) has been determined in 
\cite{grass2005}.  
We analyzed the results of our experiments from concentrated i. c.,
using for  $\langle N_{occ} \rangle (t)$\/ the same 
time dependence of eq.(\ref{eq:LogLogForm}),
founding the power-law exponent $\delta\simeq 0.54$,
but we could not establish any dependence on $r$.
Note that also starting from uniform i. c.
the evolution can be roughly divided into three stages:
a first ``crossover time'' stage, in which  $N_{occ}(t)\simeq $const. ;
the power-law stage, with superimposed log-log oscillations; the final
saturation stage, in which $N_{occ}(t)$\/ equals the fractal support value,
differently to the CIC experiments.

\begin{figure}
 \begin{center}
 
 \includegraphics[scale=0.3]{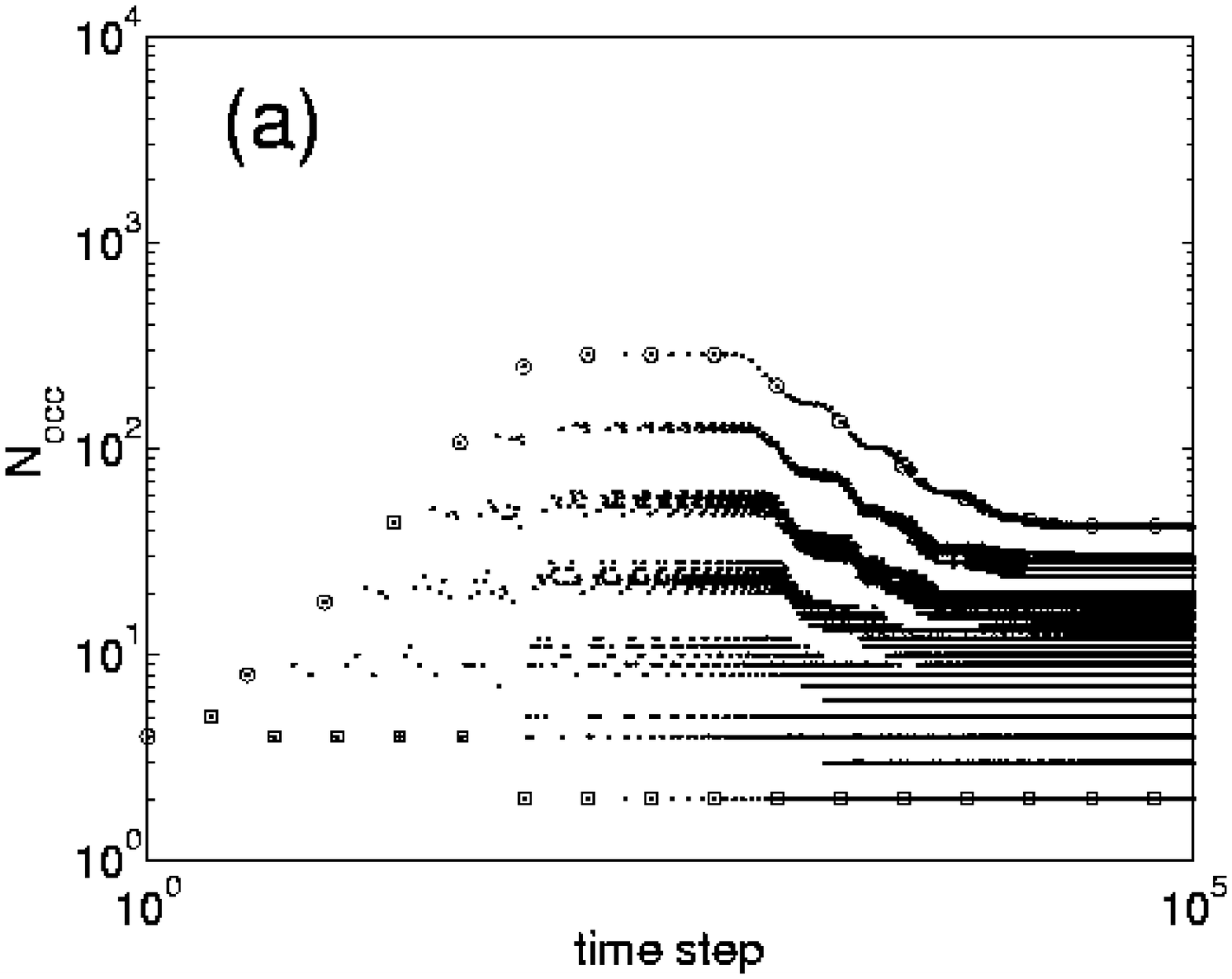}
 \includegraphics[scale=0.3]{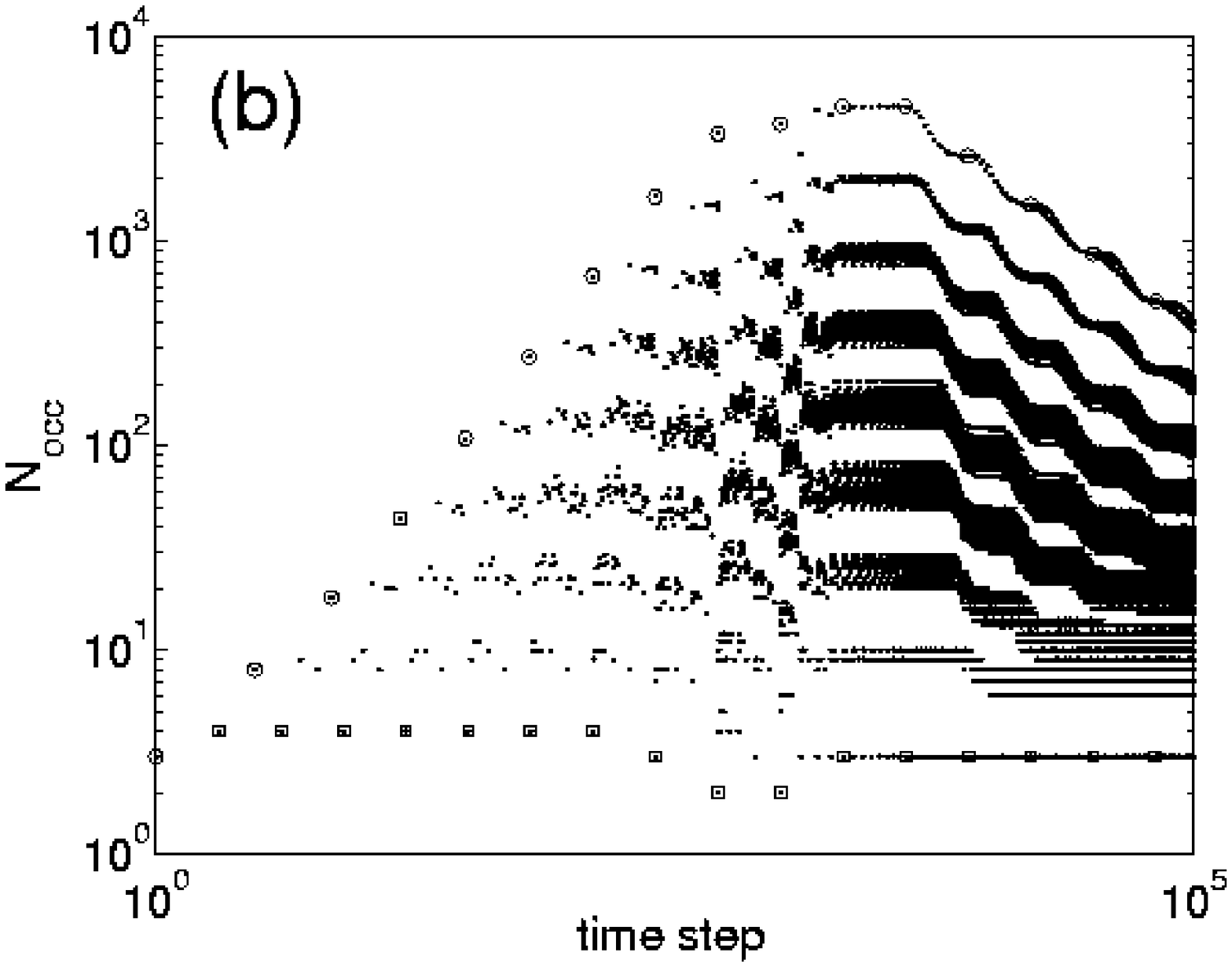}
 
 \end{center}
 \caption{\footnotesize Time evolution of the Logistic map at the critical 
          point $\mu=1.401155189$,
          the maximum values are reached at time $n=2^{k}-1$\/ (with $k=1,2,3, ...$)
          and are stressed by circles, minimum in $2^k$\/ by squares.                     
          $N_{occ}$\/ vs. time obtained by
          an expansion experiment starting from an initial condition concentrated in the 
          extremely right cell (centered in $x_c=1 - 1/(2 W_{box})$), sampling ratio $r=1$,
          number of partition cells  $W_{box}=10^5$\/ in {\bf (a)} and $W_{box}=10^7$\/ 
          in {\bf (b)}.}
 \label{fig:SingleCellEvolFig}
 \end{figure}

To investigate the origin of the observed behavior (on average) 
starting from CIC, we performed others expansion experiments, 
without averaging, selecting a special initial cell.
In fig.\ref{fig:SingleCellEvolFig} we show the results of an
expansion experiment from the $W_{box}$-th cell, that has the point
$x=1$\/ as right extreme, where 
the distance of $x=1$\/ from the Feigenbaum attractor is infinitesimal.
Again there are three evolution stages.
The first stage presents, for both $N_{occ}(t)$\/ and $S_q$, exactly
the same pattern already observed for the sensitivity function starting from $x_0=1$,
characterized by large fluctuations \cite{mori}. 
The upper limit values  correspond to 
times $n=2^k -1$\/ and form a straight line for 
$N_{occ}(t)$\/ and $S_{q=0.2445}$\/ 
as already found for the sensitivity function 
\cite{Ba:02}.
This stage is longer when $W_{box}$\/ increases.
In the second stage, the upper limit of the 
large fluctuation pattern decreases, following approximately a
power-law with superimposed log-log oscillations.
Finally a saturation stage is reached,
in which the evolution shows a regular periodic pattern.
Some features of the evolution from this special cell 
remind the averaged behavior. This is not trivial
because there are single cells which exhibit a completely different
time evolution. We checked, for instance, the repellor
$x=x_r=0.5602326...$\/ (solution of the equation $x_r = 1-\mu_{\infty} x_r^2$).
Centering the initial cell in $x_r$,
we observed a first stage in which $N_{occ}$\/ grows exponentially and
reaches its maximum  at $N_{occ} \sim W_{box}$; then it 
decreases following a power-law without log-log oscillations
and its evolution ends at a saturation value equal to 
the fractal support. 
Large fluctuations of $N_{occ}$\/ and $S_q$\/ 
do not appear at any time.

\begin{figure}
 \begin{center}

 \includegraphics[scale=0.6]{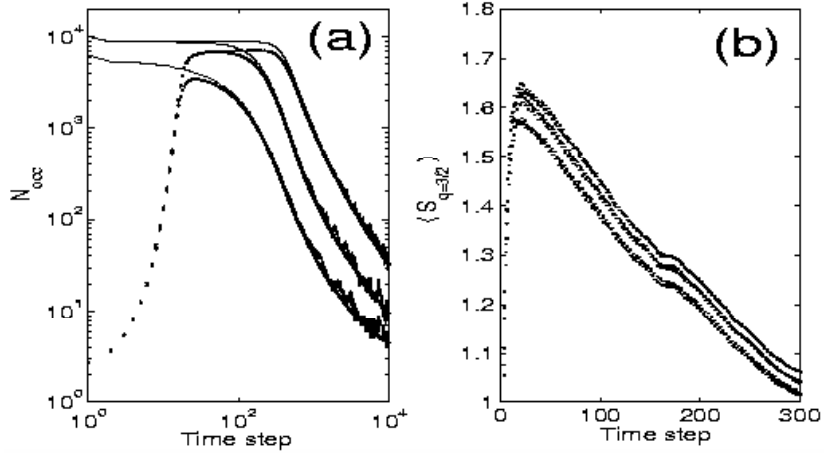}
 
 \end{center}
\caption{\footnotesize Logistic map at the tangent point $\mu=\mu_t$.
 $W_{box}=10^4$, averages performed using $N_{av}$=800 
 random choices for the initial cell.
 {\bf (a)}:  $N_{occ}$\/ from a relaxation experiment starting from uniform i. c.
 (continuous lines) and $\langle N_{occ} \rangle $\/ obtained 
 starting by a CIC (points),
 are plotted together for different $r$\/ values (from top to bottom $r=1,10,100$).
 {\bf (b)}:  Tsallis entropy with index $q=3/2$\/ is showed for different $r$\/
 (top to bottom $r=1,10,100$); the three series of points are inside 
 their statistical
 errors
 (not showed in figure).}

 \label{fig:TgPoFig}
 \end{figure}

The Feigenbaum-scaling cascade of period doubling  is not the only possible
scenario. In the following we show few comparisons with others
routes to chaos.
The logistic map at the beginning of the period three window
($\mu = \mu_t = 7/4$) exhibit both a vanishing Lyapunov exponent
($\lambda =0$) and a power-law decreasing sensitivity function
(weak insensitivity). The power-law exponent and the correspondent $S_q$\/
entropic index ($q = 3/2$) have been analytically determined
in \cite{Ba:02_2}. 
We performed for the tangent point $\mu = \mu_t$\/ the same relaxation experiments 
described in the previous part for $\mu = \mu_{\infty}$,
reproducing and improving the results already obtained in 
\cite{Co:04} and \cite{To:04}. 
Our results are showed in fig.\ref{fig:TgPoFig}.
As already stressed in \cite{Co:04}, the evolution starting from CIC
begins miming a chaotic behavior:  $S_{BG}$\/ grows linearly
and $\langle N_{occ} \rangle $\/ exponentially; then a maximum is reached.
Thereafter $\langle N_{occ} \rangle $\/ presents a ``crossover'' stage, with 
$\langle N_{occ} \rangle = const. \sim W_{box} $\/, which is longer for larger $r$. 
This has been already analyzed in \cite{To:04} for uniform i. c. but 
not for CIC. In the third stage $\langle N_{occ} \rangle $\/ decreases 
with a power-law and no log-log oscillations. The evolution ends when 
$\langle N_{occ} \rangle $\/ reaches the three points attractor. 
The curves relatives to CIC and to uniform i. c. join each other after 
the first stage (fig. \ref{fig:TgPoFig} -a), differently from the 
$\mu = \mu_{\infty}$\/ case (fig.\ref{fig:NoccCriticPoFig} -b).
Turning our attention to fig. \ref{fig:TgPoFig}-b $S_{q}$, after 
the mimed ``chaotic'' stage, decreases and is linear only for $q$ = 3/2. 
There is no crossover time for any sampling ratio and for $q$ = 3/2 
the slope depends on $W_{box}$ \cite{Co:04} but not on $r$.

\begin{figure}
 \begin{center}

 \includegraphics[scale=0.5 , angle=0]{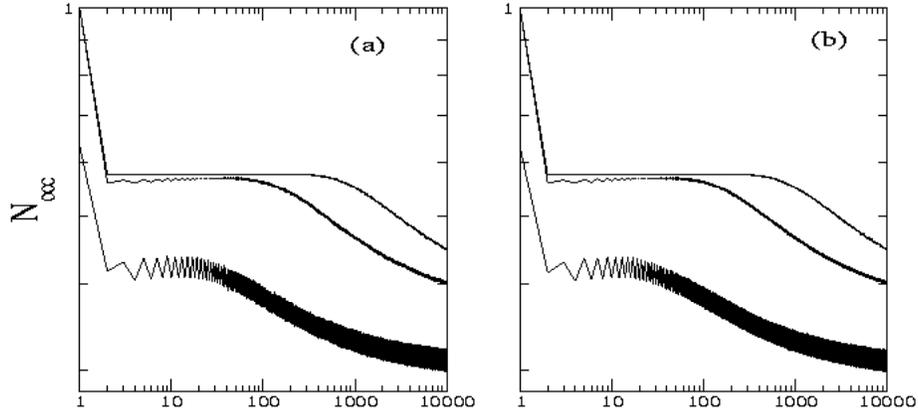}
 
 \end{center}
 \caption{\footnotesize Evolution from spread i. c. at $\mu_{1/2}$ in the 
 generalized logistic with inflection 1/2. The curves are obtained, from top to bottom, 
 using $r$ = 100, 10, 1 for $W_{box}$ = 64.000 (a) and $W_{box}$ = 128.000 (b).}

 \label{fig5}
 \end{figure}

Finally we examined the transition to chaos in the generalized logistic map
when the inflection is 1/2. Here the main bifurcation cascade, as well as
the bifurcations generated by the periodic windows in the chaotic region,
collapse in a single point \cite{inflection}. In such a point the map
undergoes the transition to chaos and the Lyapunov exponent is zero.
The experiments performed starting from uniformly spread i. c. show again
two features already encountered in the previous cases. There exists
a ``crossover time '' (fig.\ref{fig5}) which is more extended for larger $r$.
Thereafter it appears a regime of convergence to the
attractor in which N$_{occ}$ follows a power-law with negative exponent and
no log-log oscillations.
In the limits of our numerical experiments such exponent depends
neither on the grid nor on the sampling ratio.
 
\section{Conclusions}

Our numerical experiments show that the long time evolution 
on the Feigenbaum attractor from  CIC  happens in three stages,
instead of the two observed in the chaotic case.
The saturation is reached only after a power law decreasing stage,
with superimposed log-log oscillations, observed for high enough $r$\/ values.

We also showed that 
crossover-time appears in the time evolution from uniform i. c.,  
in all the cases examined. 
The crossover-time appears also in the time evolution from CIC 
of the logistic map for $\mu = \mu_t$\/
but not for $\mu = \mu_{\infty}$.

Observing $S_q$\/ in the expansion experiments from CIC for high
enough values of the entropic index $q$, log-log oscillations (for $\mu = \mu_{\infty}$)
and crossover-time (for $\mu = \mu_t$) do not appear. 
The escort distribution, implicit in the Tsallis entropy formulation eq.(\ref{eq:TsEnt}),
selects, increasing $q$, the more populated regions of the phase space. 
Thus the two observed behaviors probably originate from 
the contribution of the less populated cells.

\end{document}